\documentclass{aa}
\usepackage{graphicx}
\usepackage{txfonts}
\begin{document}
%
   \title{$uvby-H_{\beta}$ CCD Photometry of NGC~1817 and NGC~1807 
\thanks{
Tables~4 and 5 are only available in electronic form from CDS via
anonymous ftp to cdsarc.u-strasbg.fr (130.79.128.5) or 
via http://cdsweb.u-strasbg.fr/cgi-bin/qcat?J/A+A/
Fig~1 and Table~1 and 2 are only available in electronic form
via http://www.edpsciencies.org} 
    }

   \author{L. Ba\-la\-guer-\-N\'u\-\~nez\inst{1,2,3}, 
   C. Jor\-di\inst{1,4}, 
   D. Galad\'{\i}-Enr\'{\i}quez\inst{5},
   E. Masana\inst{1}
          }

   \offprints{Balaguer-N\'u\~nez, L., \email{Lola.Balaguer@am.ub.es}}

   \institute{Departament d'Astronomia i Meteorologia, Universitat de
            Barcelona, Avda. Diagonal 647, E-08028 Barcelona, Spain 
         \and
            Shanghai Astronomical Observatory, CAS Shanghai 200030,
            P.R. China
         \and
            Institute of Astronomy, 
	    Madingley Road, CB3 OHA Cambridge, UK
         \and
            CER for Astrophysics, Particle Physics and Cosmology,
	    associated with Instituto de Ciencias del Espacio-CSIC
        \and
            Centro de Astrobiolog\'{\i}a (CSIC-INTA). 
            Carretera de Ajalvir, km 4, E-28850 
            Torrej\'on de Ardoz, Madrid, Spain
             }
\date{Received ; accepted}

\authorrunning{Balaguer-N\'u\~nez et al}
\titlerunning{ $uvby-H_{\beta}$ Photometry of NGC 1817}

\abstract{
   We have investigated the area of two NGC entries, NGC~1817 and NGC~1807,
with deep CCD photometry in the $uvby-H_{\beta}$ intermediate-band system.
   The photometric analysis of a selected sample of stars of the open
cluster NGC~1817 yields a
  reddening value of $E(b-y)$ = 0.19$\pm$0.05, a distance modulus of
 $V_0-M_V$ = 10.9$\pm$0.6, a metallicity of [Fe/H] = $-$0.34$\pm$0.26 
and an age of $\log t$ = 9.05$\pm$0.05.
Our measurements allow us to confirm that NGC~1807 is 
not a physical cluster. 
   }

   \maketitle

%

\section{Introduction}

 The open cluster NGC~1817 (C0509+166), in Taurus 
[$\alpha_{2000}$=5$^{\mathrm h}12^{\mathrm m}\llap{.}1$,
$\delta_{2000}=+16{\degr}42\arcmin$], 
is an old and rich but poorly studied open cluster (Friel \cite{Friel}).
        NGC~1817 seems to be as old as the Hyades,
although comparison of the red giant clump with that of Hyades/Praesepe
suggests that NGC~1817 has a lower heavy-element abundance. Its location at
1800 pc almost directly towards the Galactic anti-centre and 400 pc below the
plane [$l=186^\circ\llap{.}13$, $b=-13^\circ\llap{.}12$], and its 
metallicity, lower than solar, make it an object of special interest for
the research of the structure and chemical evolution of the Galaxy
(Salaris et al.\ \cite{Sal}, Chen et al.\ \cite{Chen} and references 
therein).
Other clusters (NGC~2266, NGC~5822) also display a metallicity lower
than the mean [Fe/H] at their radius, reaffirming the idea that there
exists an intrinsic dispersion in [Fe/H] at any radial distance from the
Galactic centre.

Cuffey (\cite{cuff}) 
obtained extensive photographic photometry of stars in this area in the 
blue and red bands down to a limiting magnitude of $R$=14. 
Then Purgathofer (\cite{pur}, \cite{Purga}) performed a  
photometric study of the region, reaching  $V$=14. 
Harris 
\& Harris ({\cite{Har}, hereafter HH77) obtained $UBV$ photographic photometry 
of 265 stars in the central area of this cluster down to a limit $V$=16.7. 
Grocholski \& Sarajedini (\cite{Groch}) used the K-band from 2MASS photometry
as a distance indicator and unpublished $BVIK$ to compare different
theoretical isochrones (Grocholski \& Sarajedini \cite{Gro03}).

Recently, Mermilliod et al.\ (\cite{Mermi}, hereafter Mer03) 
showed photometry and radial velocity results for 88 red giant 
stars in the area, 
finding 39 members of NGC~1817 out to a distance of 27$\arcmin$. 
A radius of at least twice as large as previously 
tabulated is then expected for NGC~1817.


   NGC~1807, C0507+164, also in Taurus [$\alpha_{2000}$=5$^{\mathrm h}10^{\mathrm m}\llap{.}7$,
$\delta_{2000}=+16{\degr}32\arcmin$] shows up as a group of bright stars 
on a mildly populated background,
located close to NGC~1817.
The status of NGC~1807 is debatable but it still appears 
listed as open cluster 
(Rapaport et al.\ \cite{Rapa}). Some authors
do not consider it a physical open cluster (Becker \& Fenkart \cite{beck},
Purgathofer \cite{pur}), while others have proposed that it could constitute
a multiple system with NGC~1817 (Barkhatova \cite{bark}).

Balaguer-N\'u\~nez et al.\ (\cite{Bai}, hereafter Paper~I) performed a deep 
study of the astrometry of the 
NGC~1817/NGC~1807 area using plates with a time baseline of 81 years. 
This first analysis of the astrometric data gave as a result an 
unusually large size of the open cluster
NGC~1817 and a very poor NGC~1807.
We decided to undertake a wide-field photometric study of the whole area, 
enhancing that way the astrometric-only membership analysis, and
to derive the physical properties of the existing clusters.  
In parallel, a new membership segregation was planned based on there being only
one very extended cluster in the area. 
A recalculation of absolute proper motions, with the Tycho-2 
Catalogue as reference, has been performed
and new membership probabilities, using parametric and non-parametric approaches
have been derived (Balaguer-N\'u\~nez et al.\ \cite{Bal}, hereafter 
Paper~II).
 

    In this paper we discuss the results of the CCD photometric study 
of the area of NGC~1817 and NGC~1807, 
covering 65$^{\prime}$$\times$40$^{\prime}$ down to $V$=22. 
     Section~2 contains the details of the CCD observations and their 
transformation to the standard system. In Sect.~3 we discuss
the colour-magnitude diagrams and the process of identifying 
the sample of candidate 
cluster members. Section~4 contains the derivation of the fundamental cluster
parameters of reddening, distance, metallicity and age.
In Sect.~5 we discuss the evidence that NGC~1807 is not a real physical open 
cluster, and, finally, 
Sect.~6 summarizes our
conclusions. 


\section{The Data}

\subsection{Observations}

The photometric data were obtained in several observational runs at
Calar Alto Observatory (Almer\'{\i}a, Spain) and at Observatorio del Roque 
de los Muchachos (ORM, La Palma, Canary Islands, Spain).

        Deep Str\"omgren CCD photometry of the area was performed at Calar 
Alto in January 1999 and January 2000 using the 
1.23 m telescope of Centro Astron\'omico Hispano-Alem\'an (CAHA)  
and in February 1999 and January 2000 using the 1.52 m telescope of 
Observatorio Astron\'omico Nacional (OAN).
Further data were obtained at ORM in February 2000 using the 2.5 m 
Isaac Newton Telescope (INT) of ING (equipped with the Wide-Field Camera, WFC), 
and in December 1998 and February 2000 using the 1 m Jakobus Kapteyn 
Telescope (JKT) of ING, with the $H_{\beta}$ filter.

        The poor quality of the images obtained on the 1998/99 runs and in the OAN
2000 observations, due to adverse meteorological conditions, prevented us from making use 
of the data collected during those nights.
        A log of the observations, the total number of frames,
exposure times, seeing conditions and chip specifications, is given
in Tables~\ref{log} and \ref{chips}. The frames form a mosaic that covers
the area shown in the finding chart of the cluster (Fig.~\ref{map}). 

        We obtained photometry for a total of 7842 stars in an area of 
65$^{\prime}$$\times$40$^{\prime}$ around NGC~1817 and NGC~1807, down to a 
limiting magnitude $V$=22. 
	Due to the lack of $H_{\beta}$ filter at the WFC-INT, it was only 
possible to measure it at the JKT and CAHA telescopes, thus limiting the spatial
coverage of the mosaic with this filter.

	  Beside long, deep exposures, additional shorter exposures
were obtained in order to avoid saturation of the brightest stars.

\subsection{Data Reduction}

\subsubsection{CAHA and JKT Data}

Our general procedure has been to routinely obtain twilight sky flats for all 
the filters 
and a sizeable sample of bias frames (around 10) before and/or after every run. 
Flat fields are typically fewer in number, from five to ten per filter.  
Two or three dark frames of 2000 s were also taken. 
IRAF\footnote{IRAF is distributed by the National Optical Astronomy 
Observatories,
which are operated by the Association of Universities for Research
in Astronomy, Inc., under cooperative agreement with the National
Science Foundation.}
routines were used for the reduction process as described below.  
Dome flats were taken to check shutter effects on the images
of the OAN and CAHA telescopes where significant effects were previously noted 
(Galad\'{\i}-Enr\'{\i}quez et al.\ \cite{Gala94}, Jordi et al.\ \cite{Jordi95}).
In the current configuration of all 
the telescopes used this effect is negligible.    

The bias level was evaluated individually for each frame by averaging the counts
of the most stable pixels in the overscan areas. The 2-D structure of the bias
current was evaluated from the average of a number of dark frames with zero exposure time.
Dark current was found to be negligible in all the cases.
Flatfielding was performed using 
sigma clipped, median stacked, dithered twilight flats.

Our fields are not crowded. Thus, synthetic aperture
techniques provide the most efficient measurements of relative fluxes within 
the frames and from frame to frame. We use the appropiate IRAF packages, 
and DAOPHOT and DAOGROW algorithms (Stetson \cite{Stet87}, \cite{Stet90}). 
We analyzed the magnitude growth curves and determined the aperture correction 
with the IRAF routine MKAPFILE.

From a large number of frames with different FWHM, we applied an iterative
procedure to obtain the instrumental magnitudes. In the first step, a 
preliminary value of the FWHM was used to detect the stars recorded in each frame.
DAOPHOT and MKAPFILE were then used to obtain the instrumental photometry and 
the individual FWHM of each frame. In the second step, the detection of the
stars was improved by using the individual FWHM given by DAOGROW. Again, 
DAOPHOT and MKAPFILE were used to obtain more accurate instrumental photometry
and new individual FWHM. The iteration finishes when no new stars are detected
and when the individual FWHM are the same as in the previous step. Two iterations
were enough in our case. 

        The same field was measured with both long and short exposure. 
Cross identification of stars 
among different frames was performed using the DAOMATCH and DAOMASTER programs 
(Stetson \cite{Stet93}). 
We retained only those stars
detected in at least three filters, to enable the computation of two 
independent colours. 

Equatorial coordinates were computed using the USNO2 Catalogue 
(Monet et al.\ \cite{USNO2}) 
as reference stars. The region under study contains 2877 stars from this 
catalog. Following Galad\'{\i}-Enr\'{\i}quez et al.\ (\cite{gala1}) 
the best fitting of these reference stars was a second order pair of equations. 

\subsubsection{WFC-INT Data}

        After processing the WFC-INT frames as above, 
we found problems in the determination of the photometric 
zero point calibration for the 4 chips of the mosaic, basically 
due to gain differences between the A2D converters of the 4 CCDs. 
We decided to employ the pipeline specifically 
developed by the Cambridge Astronomical Survey Unit 
for WFC images from the INT, where the four chips are normalized 
to a common system in their level counts. The pipeline linearizes,
bias subtracts, gain corrects and flatfields the images. Catalogues
are generated using algorithms described in Irwin (\cite{Irw}).
The pipeline gives accurate positions 
in right ascension and declination linked to the USNO2 Catalogue 
(Monet et al.\ \cite{USNO2}), and instrumental magnitudes with 
their corresponding errors. A complete description can be found 
in Irwin \& Lewis (\cite{Irwin})
and in http://www.ast.cam.ac.uk/\~{}wfcsur/index.php. 

\subsubsection{Transformation to standard system}
 
Once the instrumental magnitudes and their errors were obtained, the next step
was their transformation into the standard system.

        The coefficients of the transformation equations were computed 
by a least squares method using the instrumental magnitudes of the
standard stars and the standards magnitudes and colours in the 
$uvby-H_{\beta}$ system. Up to 68 standard stars from the cluster 
M~67 (Nissen et al.\ \cite{Nissen}) were observed depending on 
the size of the field. Four to six short exposures in every 
filter were taken every night with a magnitude limit of $V$=18. 
Those standard stars with residuals greater than 
2$\sigma$ were rejected. Following Jordi et al.\ (\cite{Jordi95}),
the reduction was performed for each night independently and in two steps.

        The first step is to determine the extinction coefficients for
each passband from the standard stars. To calculate these coefficients
we make use of all the stars ($\approx$300) in the field of M67, 
increasing the accuracy
of our fit. The extinction coefficients for each night and filter are 
then calculated as the difference between the measurements as a function 
of the difference in airmasses. For instance, we have:
 $u_i - u_j = k_u ({\chi}_i - {\chi}_j)$ 
where $u_i$,$u_j$ are the different measurements
of the same star at different airmasses $\chi_i$,$\chi_j$. 
Analogous equations were used for the other passbands. 
Typical residuals are $\sigma$=0.008 for the CAHA and JKT and $\sigma$=0.001
for the WFC. 

        Because of the long exposure times, the suitable airmass value for each 
frame was obtained by integrating the instantaneous airmass throughout the 
exposure. Following Jordi et al.\ (\cite{Jordi95}) we approximated the 
integral by Simpson's rule with three points.  
        
        With the extinction coefficients fixed, the transformation to the
standard system was completed in the next step, in which we used only the 
stars with known standard photometric values present in the field. The equations were as follows:

        $y'$- $V$ = $a_1 + a_2 (b-y)$

        $(b-y)'$ = $a_3 + a_4 (b-y)$

        $c_1'$ = $a_5 + a_6 (b-y) + a_7 c_1$

        $m_1'$ = $a_8 + a_9 (b-y) + a_{10} m_1$

        $H_{\beta}'$ = $a_{11} + a_{12} H_{\beta}$

	\noindent were $a_i$ are the transformation coefficients and 
the $'$ indicates instrumental values.

        In the CAHA images we decided to treat all the nights together 
---after correcting for atmospheric extinction--- to fit those coefficients, 
and then to determine the zero point deviations from that fit for every night. 
For coherence, we refer those zero points to the precise results obtained 
with the WFC-INT, thanks to the generous overlap between the CAHA observations
and the WFC-INT field.

The internal errors of the individual measurements were computed as described 
by Jordi et al.\ (\cite{Jordi95}), taking into account the errors in the
instrumental magnitudes on the one hand, and the errors in the transformation 
equations on the other hand. Final magnitudes, colours and errors were obtained by
averaging the individual measurements of each star using the internal error
for weighting (Galad\'{\i}-Enr\'{\i}quez et al.\ \cite{gala1}, Rossell\'o et al.\
\cite{Ross85}). The final errors as a function of apparent visual magnitude
are given in Table~\ref{error} and plotted in Fig.~\ref{errby}. The
structure in the magnitude dependence is owed to the mosaic of images 
from different nights and different telescopes having different limiting 
magnitude. 

\addtocounter{table}{2}

\begin{table*}
\caption { Number of stars observed ($N$) and mean internal errors 
           ($\sigma$) as a function of apparent visual magnitude.}
\begin {tabular} {ccccccccccc}
\hline
\hline
 $V range$ & $V$ && $(b-y)$ && $m_1$ && $c_1$ && $H_{\beta}$ \\
\hline
 & $N$&$\sigma$ & $N$&$\sigma$ & $N$&$\sigma$ & $N$&$\sigma$ &$N$&$\sigma$ \\
\hline
 8- 9 &    3 & 0.011 &    3 & 0.068 &    3 & 0.012 &    3 & 0.042 &    2 & 0.007 \\
 9-10 &    8 & 0.010 &    8 & 0.030 &    8 & 0.041 &    8 & 0.043 &    4 & 0.014 \\
10-11 &   18 & 0.008 &   18 & 0.007 &   16 & 0.014 &   14 & 0.014 &    8 & 0.039 \\
11-12 &   36 & 0.010 &   36 & 0.013 &   33 & 0.016 &   32 & 0.022 &   19 & 0.026 \\
12-13 &  102 & 0.011 &  102 & 0.014 &  100 & 0.025 &   95 & 0.026 &   65 & 0.018 \\
13-14 &  167 & 0.009 &  167 & 0.011 &  165 & 0.023 &  158 & 0.025 &   86 & 0.018 \\
14-15 &  300 & 0.009 &  299 & 0.014 &  297 & 0.019 &  292 & 0.022 &  151 & 0.028 \\
15-16 &  471 & 0.009 &  471 & 0.016 &  467 & 0.024 &  458 & 0.031 &  216 & 0.024 \\
16-17 &  833 & 0.014 &  833 & 0.026 &  799 & 0.038 &  761 & 0.047 &  355 & 0.027 \\
17-18 & 1110 & 0.021 & 1110 & 0.036 &  945 & 0.050 &  843 & 0.058 &  317 & 0.037 \\
18-19 & 1140 & 0.019 & 1139 & 0.038 &  882 & 0.045 &  734 & 0.055 &   36 & 0.036 \\
19-20 & 1204 & 0.012 & 1204 & 0.029 &  991 & 0.036 &  565 & 0.056 &      &       \\
20-21 & 1119 & 0.017 & 1119 & 0.037 &  867 & 0.055 &  143 & 0.072 &      &       \\
21-22 & 1007 & 0.045 & 1007 & 0.064 &  387 & 0.090 &    5 & 0.092 &      &       \\
22-23 &  252 & 0.090 &  252 & 0.121 &   32 & 0.139 &      &       &      &       \\
\hline
Total & 7770 &       & 7768 &       & 5992 &       & 4111 &       & 1259 &       \\
\hline
\end {tabular}
\label{error}
\end {table*}

\addtocounter{figure}{1}

\begin{figure}
\begin{center}
\resizebox{\hsize}{!}{\includegraphics{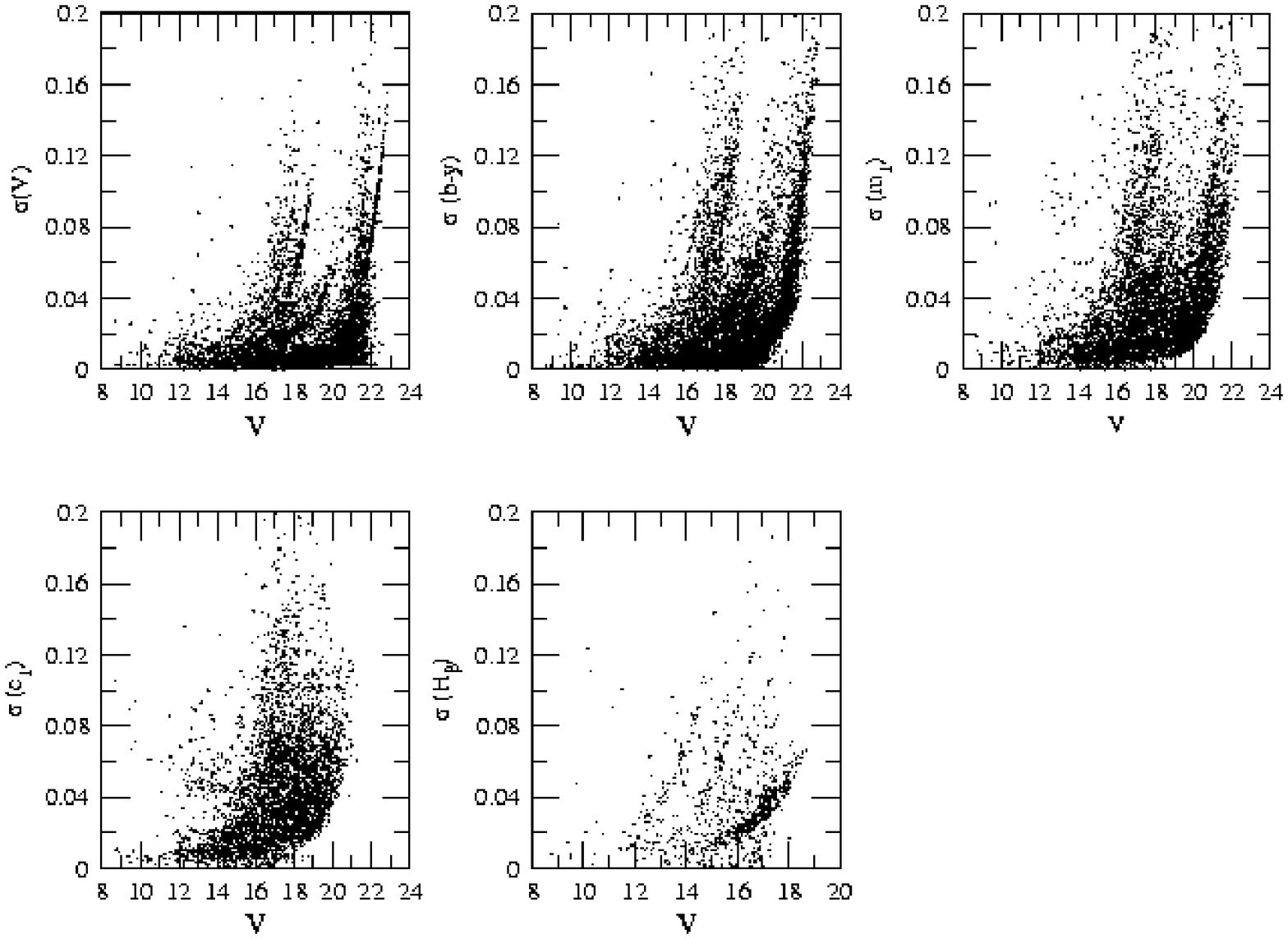}}
\end{center}
\caption{Internal errors of magnitude and colours as a function of the     
         apparent visual magnitude, $V$, 
for all stars in the cluster region. The
structure in the magnitude dependence is owed to the mosaic of images
from different nights and different telescopes having different limiting
magnitude.}
\label{errby}
\end{figure}

Table~4 lists the $u,v,b,y,H_{\beta}$ data for all 7842 stars in a
region of 65$^{\prime}$$\times$40$^{\prime}$ around the
open cluster NGC~1817 (Fig.~\ref{map}). Star centres are given as frame ($x,y$)
and equatorial ($\alpha_{\mathrm J2000}$,$\delta_{\mathrm J2000}$) coordinates. 
Given the large 
area coverage and depth of our photometry, a new numbering system 
for stars in this field is introduced. 
An identification number was assigned to each star following 
the order of increasing right ascension.
Column 1 is the ordinal star number; columns 2 and 3
are $\alpha_{\mathrm J2000}$ and $\delta_ {\mathrm J2000}$;
columns 4 and 5 are the respective $x$, $y$ coordinates in arcmin;
columns 6 and 7 are the $(b-y)$ and its error,
8 and 9 the $V$ magnitude and its error, 10 and 11 the $m_1$ and its error, 
12 and 13 the $c_1$ and its error, 
and 14 and 15 the $H_{\beta}$ and its error. 
In column 16, stars considered candidate members (Sect.~3.1.) are labelled
 'M', while those classified as non-members show the label 'NM'.

\addtocounter{table}{1}

    The cross-identification of stars in common
    with the astrometry (Paper~II), WEBDA
(htpp://obswww.unige.ch/WEBDA), 
Hipparcos (ESA, \cite{esa}), Tycho-2 (H\a{o}g et al.\ \cite{tyc2}) 
and USNO-2 (Monet et al.\ \cite{USNO2}) catalogues
    is provided in Table~5. 
\addtocounter{table}{1}

\subsection{Comparison with Previous Photometry}

	Only four stars in the field of NGC~1817 have been previously studied 
using Str\"omgren photometry, in the range of $V$=9.5 to 12.9. 
Maitzen et al.\ (\cite{Maitz}) performed a 
photometric search for Ap stars among Blue Stragglers in open clusters,  
with mean errors of less than 0.01.
Mean differences in the sense ours minus others are:
0.03($\sigma$=0.04) in $V$, 0.02 ($\sigma$=0.04) in $b-y$,
$-$0.03 ($\sigma$=0.07) in $m_1$ and 0.00 ($\sigma$=0.03) in $c_1$
for the 4 stars in common.

	On the other hand, the $V$ magnitude derived from the $y$ filter
can be compared with the published broadband data. 
	The recent study of radial velocities of red giants in the area 
by Mer03 included photoelectric photometry of 25 stars. 
We have 22 common stars with the photoelectric photometry of HH77 
and 19 common stars with Purgathofer (\cite{Purga}), up to $V$=16.
	The corresponding mean differences in $V$, in the sense ours minus others
are: $-$0.01(0.03),$-$0.01(0.04) and $-$0.01(0.05), respectively. 
Transformation between
$B-V$ and $b-y$ from several authors (see Moro \& Munari, \cite{ADPS}) fails 
to cover the whole range under study. Making use of merged broadband photometry,
we can find a linear relation between the two indices: 
$B-V$=(1.719$\pm$0.035)$(b-y)-(0.170\pm0.024)$, $N$=62. 
The standard deviation of 
the residuals about the mean relation is 0.055, where the typical 
uncertainty in $B-V$ is 0.02 and in $b-y$ is 0.014.


\section{Colour-Magnitude Diagrams}

        We use the $V vs\ (v-y)$ colour-magnitude diagram 
for our study. As Meiborn (\cite{Mei00}) stated, the colour-magnitude 
diagram based on this
colour index defines the main-sequence of a cluster significantly better 
than the traditional $V vs\ (b-y)$ diagram (Fig.~\ref{HR} left and centre).
The fact that the reddening vector runs parallel 
to the cluster sequence in this diagram
allows a much better separation of cluster and field stars. 

        The observational colour-magnitude diagram for all the stars
in the studied area 
(Fig.~\ref{HR}) displays a fairly well defined main sequence. Specially 
outstanding against the field background is the red giant clump and the 
main sequence between 0.8$<(v-y)<$1.8. 

\begin{figure*}
\resizebox{17cm}{!}{\includegraphics{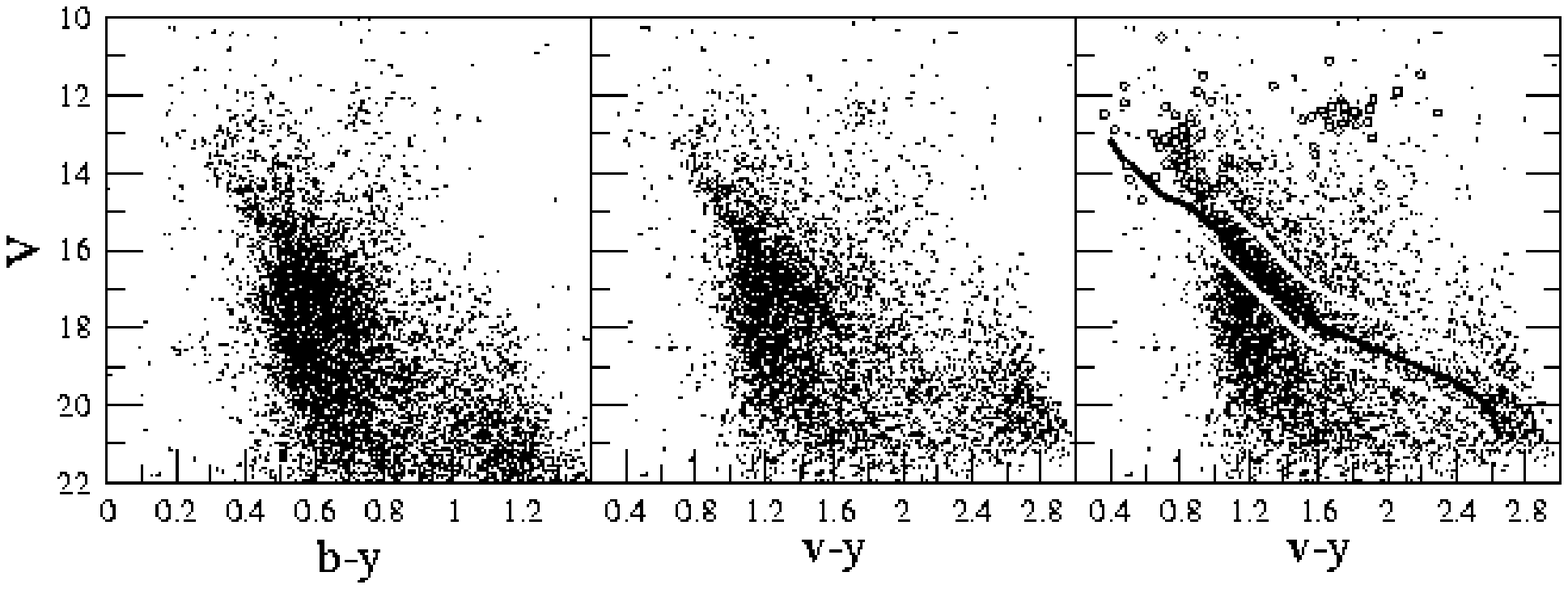}}
\caption{The colour-magnitud diagram of the NGC~1817 area.
Empty circles in the right figure are the astrometric members from Paper~II. 
Black line is a shifted ZAMS, with the chosen margin for candidate members
($V+0.5$,$V-1$) in white lines. See text for details.}
\label{HR}
\end{figure*}

\subsection{Selection of candidate member stars}

	The first step in the study of an open cluster is the determination
of its member stars. 
A selection of probable member stars 
up to the limiting magnitude of our sample
can be obtained
combining astrometric with photometric criteria. 
Unfortunately, proper motions 
(Paper~II) and radial velocities
(Mer03) 
are only available for the 
brightest stars in the area. 
Photometric 
measurements help to reduce the possible field contamination in the
proper motion membership ---among bright stars---,
as well as to enlarge the selection of members towards faint magnitudes.  

\begin{figure}
\resizebox{\hsize}{!}{\includegraphics{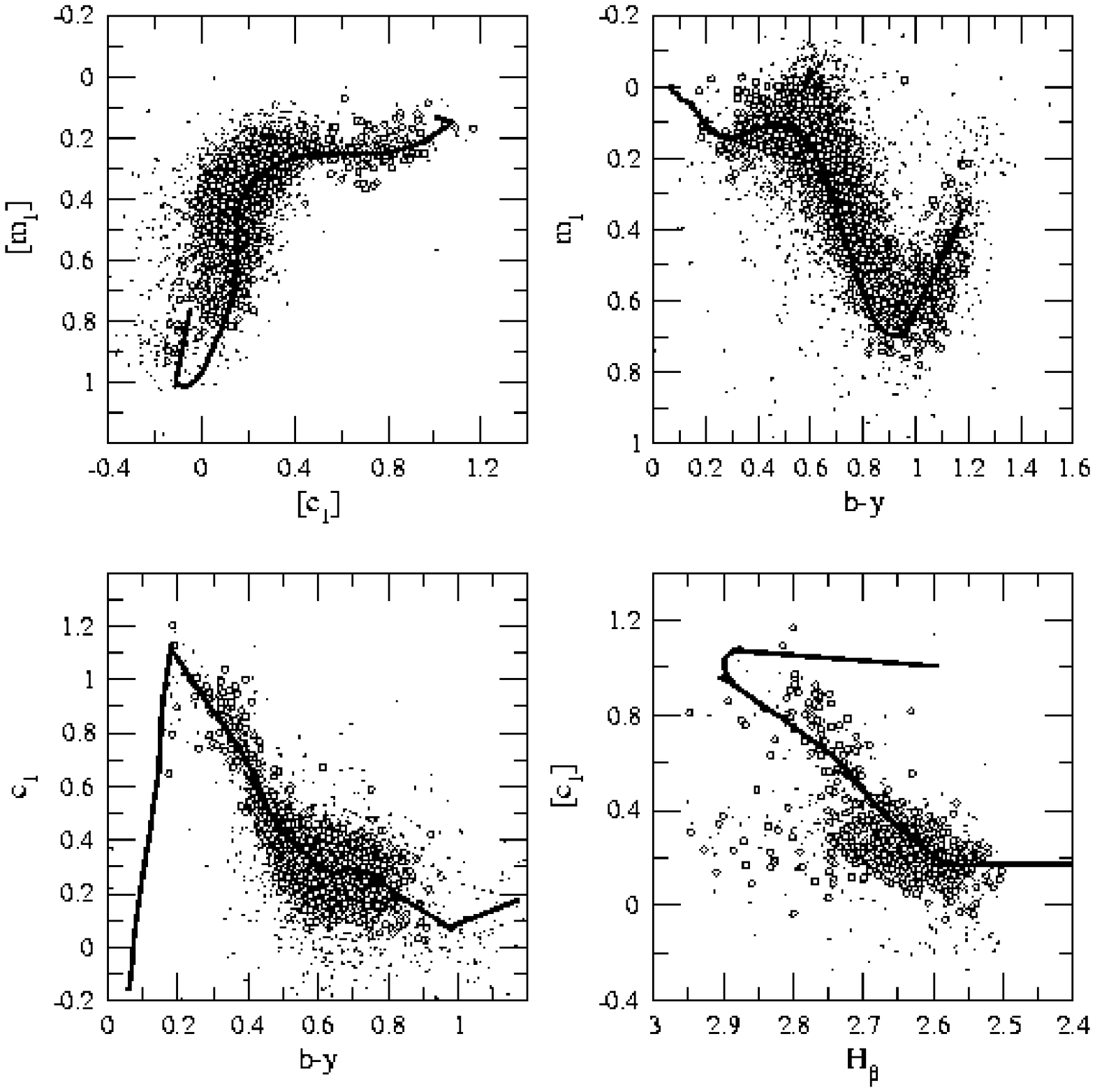}}
\caption{The colour-colour diagrams of NGC~1817.
           Empty circles denote candidate members of NGC~1817,
chosen with astrometric and non-astrometric criteria as 
explained in Sect.~3.1.
           The thick line is the standard relation shifted 
 $E(b-y)$ = 0.19 when necessary.} 
\label{colour}
\end{figure}

Among those astrometric member stars, we find 13 stars that are not 
compatible with 
the sequence of the cluster outlined in the colour-magnitude diagram.
Radial velocity information is also taken into account,
to reject stars with radial velocities incompatible with membership
and to include stars considered members by Mer03 and compatible
with membership according to our photometric data.
Our astrometric study -based on proper motion from plates-, and thus its
segregation of member stars, has a limiting magnitude of $V$=14.5. 
From this point down to our photometric limit, $V=22$ we construct a ridge line
following a fitting of the   
observational ZAMS (Crawford \cite{Craw75}, \cite{Craw78}, \cite{Craw79},
Hilditch et al.\ \cite{Hil83}, Olsen \cite{Ols84})  
in the $V - (v-y)$ diagram. 
A selection of stars based on the distance to this ridge line is then
obtained. The chosen margin for candidates includes all the stars 
between $V+0.5$ and $V-1$ from the ridge line, as shown in the
right panel of Fig.~\ref{HR}. The margins were chosen to account for
observational errors and the presence of multiple stars.

This preliminary photometric selection is refined in the colour-colour 
diagrams (Fig.~\ref{colour}) with the help of the standard relations from the
same authors.  
A final selection of 1592 stars in the area is plotted in Fig.~\ref{colour}
as empty circles in the $[m_1] - [c_1]$, $m_1 - (b-y)$, $c_1 - (b-y)$ and 
$[c_1] - H_{\beta}$ diagrams.

\section{Physical parameters of the cluster}

Narrow and intermediate passband photometry  
constitutes a useful technique for classification of the stars (Stromgren
\cite{Strom}, Philip et al.\ \cite{Phil}).
The stars of the area selected as possible cluster members were
classified into photometric regions and their physical parameters were determined. 
The algorithm uses 
$uvby-H_{\beta}$ photometry and standard relations among colour indices
for each of the photometric regions of the HR diagram. 
Masana et al.\ (\cite{Mas}) provide improvements to the algorithm 
classification and parametrization described in Jordi 
et al.\ (\cite{Jordi97}) and Figueras et al.\ (\cite{Fig}). Improvements
concern the inclusion of a grid of temperatures and gravities dependent on 
metallicity, the determination of masses and radii as well as the detection
of peculiar stars.

\begin{figure}
\resizebox{\hsize}{!}{\includegraphics{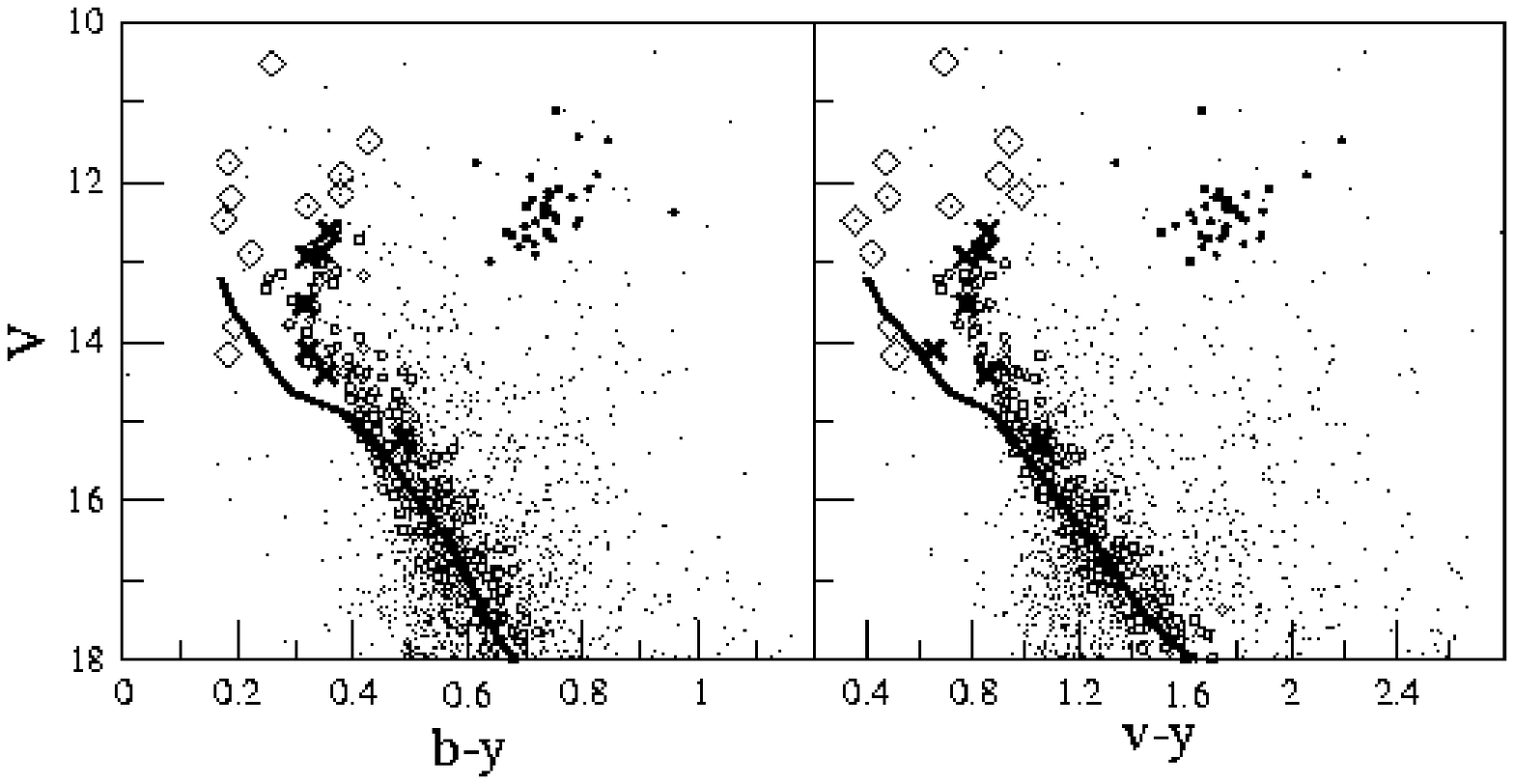}}
\caption{Colour-magnitude diagram with the subset of 264 stars used to
derive the physical parameters in empty circles, the
red giants in full circles, 
blue stragglers in diamonds and $\delta$ Scuti in crosses. Field stars are in dots.
The thick line is the ZAMS, shifted in extinction and distance modulus. 
}
\label{Vvyby1}
\end{figure}

        Absolute magnitude, effective temperature and gravity as well 
as the corresponding reddening, distance modulus, metallicity and a raw
spectral type and luminosity class are calculated 
for each star. Typical errors are 0.25~mag in $M_V$, 0.15~dex in [Fe/H],
270~K in $T_{\rm eff}$, 0.18~dex in $\log g$ and 0.015~mag in $E(b-y)$.
Even if the result for the physical parameters for a given star 
would be inaccurate ---mainly due to 
peculiarity, emission lines or binary character---,
the high proportion of "normal" stars 
makes it possible to ascertain the physical parameters of the cluster. 

\begin{figure}
\resizebox{\hsize}{!}{\includegraphics*{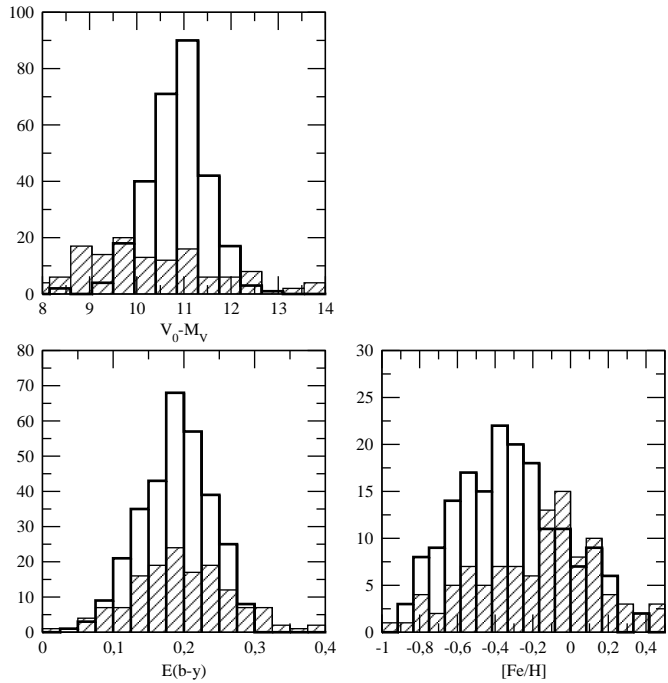}}
\caption{The histograms of the distance modulus, reddening and 
metallicity of the selected member stars of NGC~1817 (white), 
and the corresponding ones for the stars in the NGC~1807 area (dashed).}
\label{pf}
\end{figure}

	Only 525 stars among the 1592 candidate members have $H_{\beta}$
measurements. So the computation of physical parameters is only possible
for that subset. Red giant stars are outside of the validity of the 
calibrations. 
Excluding peculiar stars and those with inconsistence among their
photometric indices, we apply
an average with a 2$\sigma$ clipping to that subset and   
only 264 stars remain (Fig.~\ref{Vvyby1}).
We found a reddening value of $E(b-y)$~=~0.19$\pm$0.05
(corresponding to $E(B-V)$~=~0.27)
and a distance modulus of $V_0-M_V$~=~10.9$\pm$0.6. 
Metallicity is better calculated studying only the 196 F and G stars in 
our sample, inside the validity range of the calibrations of 
Schuster \& Nissen (\cite{Schu},
see Masana et al.\ \cite{Mas} for more details). 
We found a value of [Fe/H]~=~$-$0.34$\pm$0.26. 
        Figures~\ref{pf} show these results. 

	Our results are consistent with previous results. 
HH77 give a reddening value of $E(B-V)$~=~0.28$\pm$0.03 and a 
distance modulus of 11.3$\pm$0.3.
Friel \& Janes (\cite{Friel93}) give a value of [Fe/H]~=~$-$0.39$\pm$0.04 
($N$=4) on the basis
of moderate-resolution spectra, while Taylor (\cite{Tay}) reexamines the errors 
in these data and concludes a lower value of [Fe/H]~=~$-$0.42$\pm$0.07.  
Friel et al.\ (\cite{Friel02}) give a recalibrated value of 
[Fe/H]~=~$-$0.29$\pm$0.05 based on 3 stars. 	
Twarog et al.\ (\cite{Twa}) find an $E(B-V)$~=~0.26, an [Fe/H]~=~$-$0.27$\pm$0.02 
and $V_0-M_V$~=~11.34 based on main-sequence fitting.  
Dutra \& Bica (\cite{Dutr}) using DIRBE/IRAS 100 $\mu m$ dust emission 
integrated throughout the Galaxy derive a value of $E(B-V)_{FIR}$~=~0.33.

	The recent publication by Clem et al.\ (\cite{Clem}) 
of empirically constrained colour-temperature relations in the Str\"omgren
system allows the transformation of isochrones from the theoretical to
the observational colour-magnitude diagram. Several sets of isochrones
have been used to analyze our results.
The best fitting is
found for the Schaerer et al.\ (\cite{Scha}) isochrones. Figure~\ref{iso}
shows isochrones of Z=0.008 shifted by a reddening of 0.19 and an apparent 
distance modulus of 11.6. We found a best estimation of the
age of $\log t$ = 9.05$\pm$0.05.  
	HH77 give a Hyades-age for this cluster: 0.8 Gyr ($\log t$= 8.9). 
But a recent determination of ages of old open clusters 
(Salaris et al.\ \cite{Sal}) gives an age of 
1.12$\pm$0.18 Gyr based on the morphological age index provided by 
Janes \& Phelps (\cite{JP94}) but on a new highly homogeneous and 
reliable calibration in terms of absolute ages. 
Our age determination, 1.1~Gyr, agrees very well.

\begin{figure*}
\begin{center}
\resizebox{12cm}{!}{\includegraphics{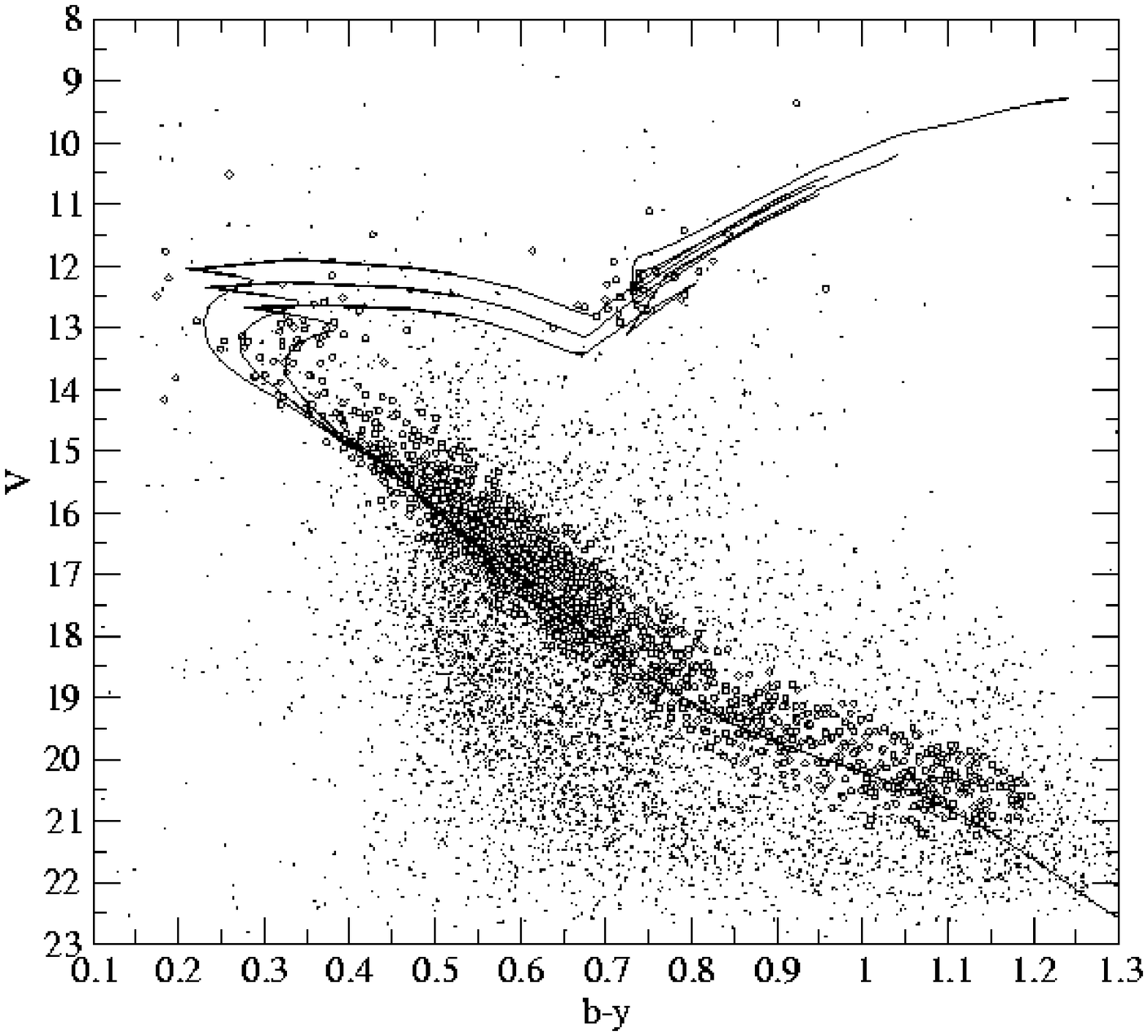}}
\caption{Geneva isochrones (Schaerer et al.\ \cite{Scha})
of $\log t$ 8.9, 9.0 and 9.1 and Z=0.008.
The adopted reddening and apparent distance modulus are 0.19 and 11.6,
respectively.
           Empty circles denote candidate members of NGC~1817.} 
\label{iso}
\end{center}
\end{figure*}

	There are seven known blue stragglers in our photometry from the 
catalogue of blue stragglers in open clusters 
(Ahumada \& Lapasset \cite{Ahu95}). But two of them (HH2090, HH1073) are not 
proper motion members. 
From a sample of 18 clusters, Wheeler (\cite{Wheel})
found a ratio of the number of stragglers to giants of
from 1/3 to 1/4. Having 39 confirmed red giants in the cluster 
(Mer03) we should expect between 10 and 13
blue stragglers.
The location in our colour-magnitude diagram of six blue stars 
that are astrometric members according to Paper~II lead us to 
classify them also as blue stragglers.
Their photometry and astrometric segregation    
from Paper~II are listed in Table~\ref{BlueS}.
 
\begin{table}
\caption {Blue Stragglers in NGC~1817, with their membership 
from Paper~II. The first seven are from the catalog of 
Ahumada \& Lapasset (\cite{Ahu95}). The last six have been classified
in this work.	 
         }
\begin {tabular} {cccccccccc}
\hline
  Id.  & HH & $(b-y)$ & $V$ &  $P$  \\
\hline
 
7091 & 2045 &  0.321$\pm$0.0059 & 12.305$\pm$0.0176 & M  \\
 157 & 2090 &  0.209$\pm$0.0097 & 13.112$\pm$0.0000 & NM \\
 389 & 2072 &  0.188$\pm$0.0011 & 12.195$\pm$0.0009 & M  \\
 387 & 2110 &  0.184$\pm$0.0051 & 11.762$\pm$0.0056 & M  \\
7467 & 2020 &  0.221$\pm$0.0126 & 12.896$\pm$0.0074 & M \\
7746 & 1073 &  0.421$\pm$0.0147 & 12.270$\pm$0.0123 & NM \\
7310 & 3001 &  0.197$\pm$0.0210 & 13.816$\pm$0.0309 & M  \\
1126 &      &  0.174$\pm$0.0027 & 12.486$\pm$0.0029 & M  \\
 390 &      &  0.379$\pm$0.0036 & 12.157$\pm$0.0048 & M  \\
 388 &      &  0.379$\pm$0.0101 & 11.914$\pm$0.0035 & M  \\
  45 &      &  0.427$\pm$0.0014 & 11.498$\pm$0.0172 & M  \\
6161 &      &  0.259$\pm$0.0023 & 10.514$\pm$0.0027 & M  \\
7332 &      &  0.183$\pm$0.1662 & 14.169$\pm$0.0965 & M  \\

\hline
\label{BlueS}
\end{tabular}
\end{table}

 	Open clusters with an age around 1 Gyr, with the turn off on the
instability strip, seem to have the largest number of pulsators. There are 
seven known $\delta$ Scuti stars in the area (Frandsen \&
Arentoft \cite{Fran98}). Five of them have absolute proper motions and 
membership probabilities from Paper~II. We have calculated the physical 
parameters for these stars and we are able to confirm the membership 
of four of them (Table~\ref{Scuti}). 

\begin{table*}
\caption {Physical parameters of the known $\delta$ Scuti stars 
in the NGC~1817 area, with the membership segregation from Paper~II. 	 
         }
\begin {tabular} {ccccccccccc}
\hline
  Id.  & $E(b-y)$& $(b-y)_0$ & $V_0$ & $M_V$ & $V_0 - M_V$ & $T_{\mathrm eff}$ & $\log g$ & $M/M_{\sun}$ & $R/R_{\sun}$ & $P$ \\
\hline

  154 & 0.179 & 0.167 &12.116 & 0.57 &11.55 &  7105 &3.22 & 2.328 & 4.613 & M  \\
  155 & 0.160 & 0.160 &12.257 & 3.95 &11.44 &  7199 &3.35 & 2.200 & 3.858 & M  \\
  167 & 0.211 & 0.106 &12.631 & 1.77 &10.86 &  7794 &3.92 & 2.035 & 2.545 & NM \\
  184 & 0.216 & 0.136 &13.494 & 3.30 &10.20 &  7731 &4.44 & 1.588 & 1.369 & -- \\
  211 & 0.227 & 0.260 &14.256 & 3.95 &10.31 &  6780 &4.72 & 1.246 & 0.972 & -- \\
 7298 & 0.168 & 0.154 &13.405 & 2.77 &10.63 &  7502 &3.79 & 1.461 & 1.121 & M  \\
 7615 & 0.158 & 0.200 &11.946 & 1.37 &10.58 &  6917 &3.44 & 2.007 & 3.376 & M  \\
\hline
\label{Scuti}
\end{tabular}
\end{table*}


\section{NGC~1807: not a real physical open cluster}

 	 Comparison of the colour-magnitude diagram of the area 
around NGC~1807 and the
centre of the open cluster NGC~1817, Fig.~\ref{comp}, already shows a lack of 
a reliable main sequence for NGC~1807.  

\begin{figure}
\resizebox{\hsize}{!}{\includegraphics*{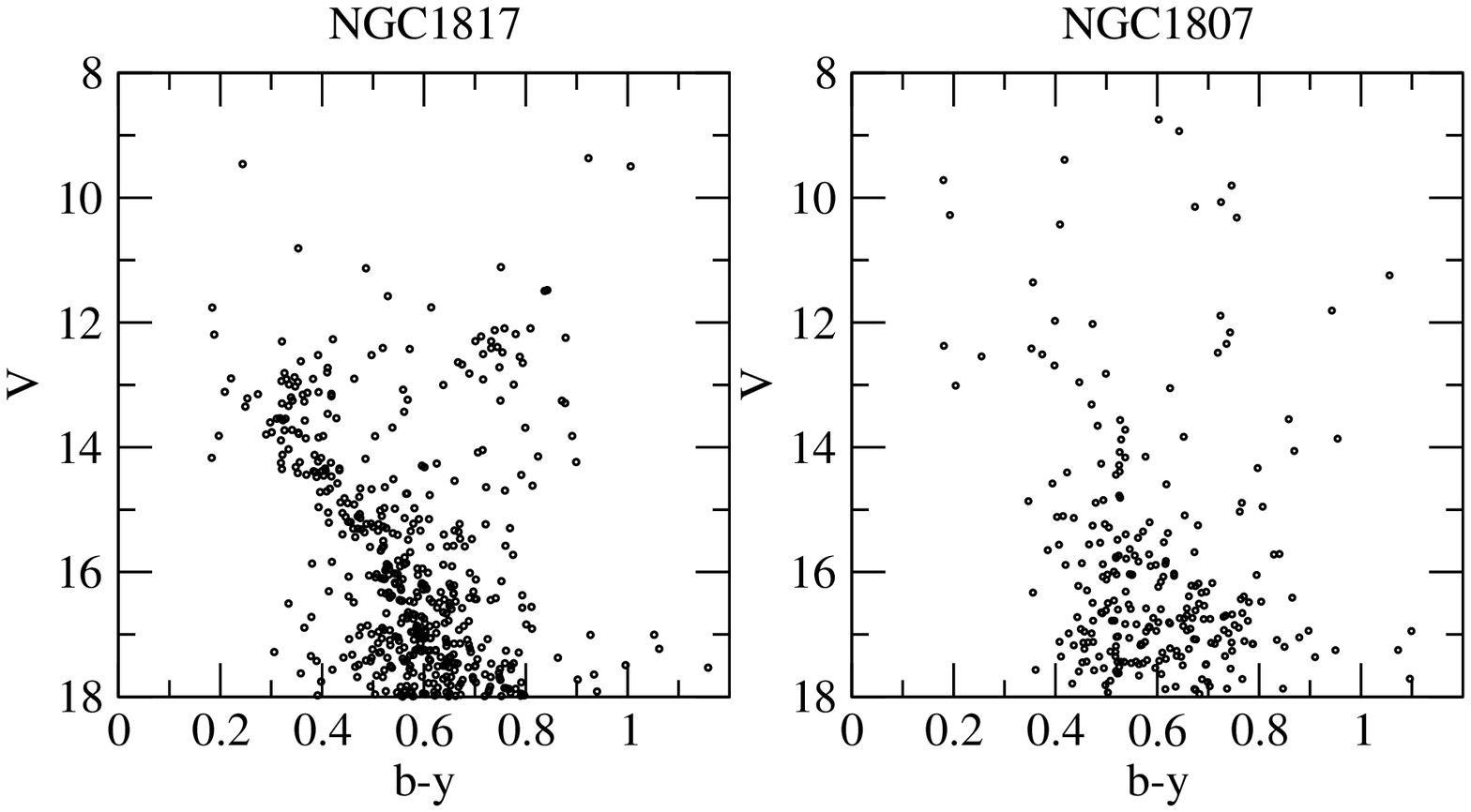}}
\caption{The colour-magnitude diagram of stars in an area of 15$\arcmin$ 
 around the centres of NGC~1817 and NGC~1807.}
\label{comp}
\end{figure}

	The star HD~33428, one of the brightest stars in the area of 
NGC~1807 ($V$=8.5), has DDO observations 
(Yoss et al. \cite{Yoss}, Piatti et al. \cite{picla}), but 
proper motion studies (Paper~II and 
Hipparcos Catalogue, ESA \cite{esa}) show that it cannot belong to 
NGC~1807. Even in the case of NGC~1807 being a real cluster, this 
star is a high proper motion star.

	The comparison of the result of our 
study of the physical parameters of NGC~1817 with a study of 200 stars 
in the area of NGC~1807 (Fig.~\ref{pf}, shaded histogram), shows 
that the lack of a clear trend is unambiguous.

Paper~II also supports the idea of a unique and very extended
cluster in the area, showing no hint of two kinematically distinct 
clusters in the area. 

	The results from Mer03 of radial velocities of red 
giants in the area also agree with there being only one 
and very extended cluster. 
There are six stars from Mer03 list in the  
NGC~1807 area. Two of them (M1152, M1208) are members of NGC~1817 and
the other four (M598, M682, M1153, M1161) have discordant radial velocities 
(+74.51, +111.53, +28.49, +11.71 km~s$^{-1}$). 

We can conclude that there is no photometric or astrometric evidence 
supporting the existence of a real cluster NGC~1807.

\section{Conclusions}

        In this paper we give a catalogue of accurate
$uvby-H_{\beta}$ and J2000 coordinates for 7842 stars in an area of
65$^{\prime}$$\times$40$^{\prime}$ around NGC~1817. 
 
	We give a selection of probable members of NGC~1817, combining this 
photometric study with the previous astrometric analysis (Paper~II). 
A better determination of the physical parameters of this extended cluster 
based on our accurate photometry gives: 
$E(b-y)$ = 0.19$\pm$0.05, [Fe/H] = $-$0.34$\pm$0.26, 
a distance modulus of $V_0-M_V$ = 10.9$\pm$0.6 and an age of 
$\log t=$9.05$\pm$0.05.
The values are consistent with
previous studies (see Hou et al.\ \cite{Hou} and references therein). 

We have not found any support for considering 
NGC~1807 a real physical cluster.     

\vspace{3mm}
%

\begin{acknowledgements}
    We would like to thank Simon Hodgkin and Mike Irwin for their 
    inestimable help in the reduction of the images taken at the WFC-INT. 
    L.B-N, also wants to thank Gerry Gilmore for his continuous help and 
    valuable comments, as well as all the people at the IoA (Cambridge)
    for a very pleasant stay. L.B-N. gratefully acknowledges financial 
    support from EARA Marie Curie Training Site (EASTARGAL) during her 
    stay at IoA.
Based on observations made with the INT and JKT telescopes operated on 
the island of La Palma by the RGO in the Spanish Observatorio del Roque 
de Los Muchachos of the Instituto de Astrof\'{\i}sica de Canarias, and  
with the 1.52~m telescope of the Observatorio Astron\'omico Nacional (OAN) 
and the 1.23~m telescope at the German-Spanish Astronomical Center,
Calar Alto, operated jointly by Max-Planck Institut f\"ur Astronomie and 
Instituto de Astrof\'{\i}sica de Andalucia (CSIC). 
    This study was also partially
    supported by the contract No. AYA2003-07736 with MCYT.
    This research has made use of Aladin, developed by CDS, 
Strasbourg, France.

\end{acknowledgements}

\Online

\setcounter{figure}{0}
\setcounter{table}{0}

\begin{figure*}
\resizebox{\hsize}{!}{\includegraphics{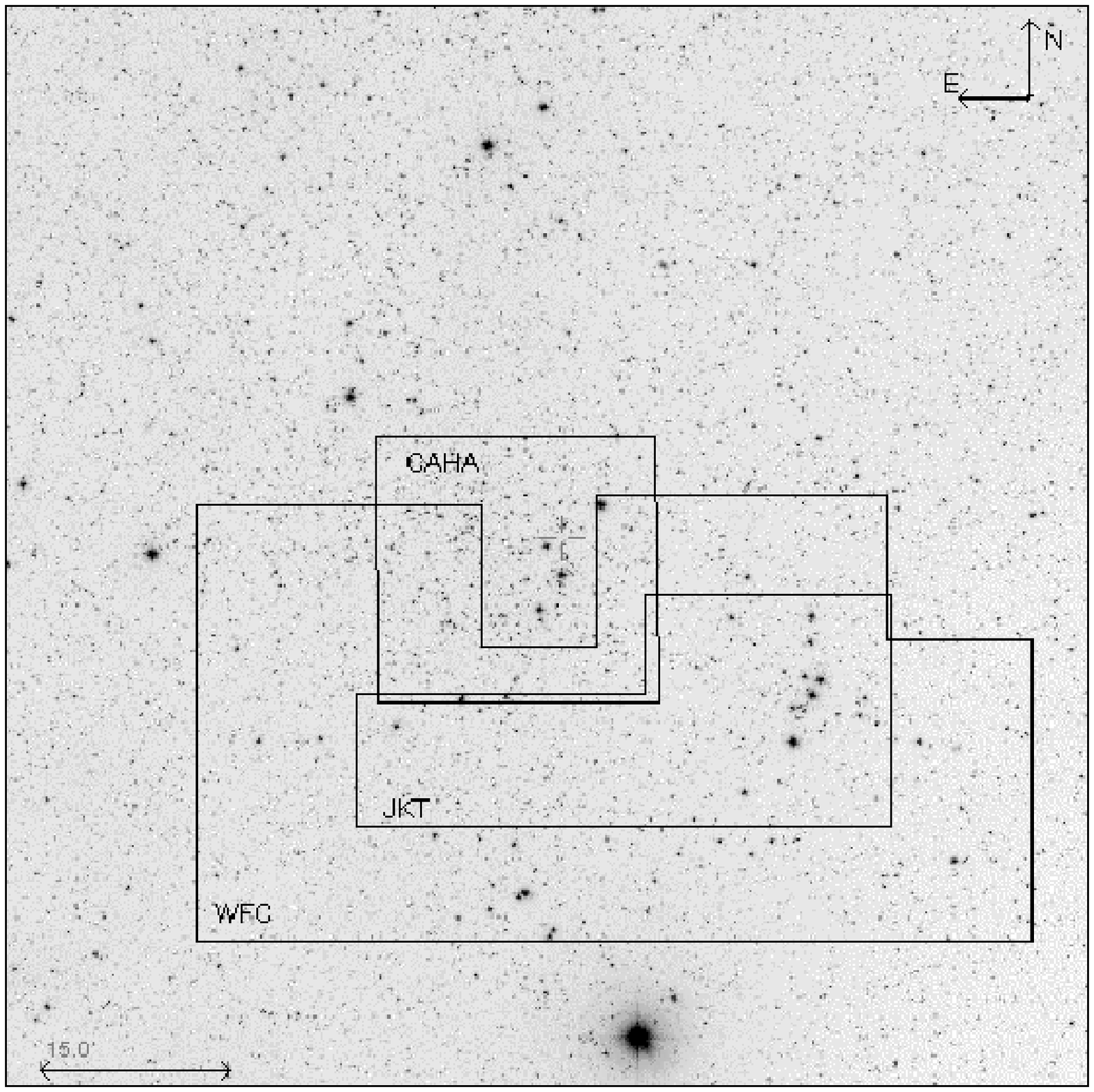}}
\caption{Finding chart of the area under study. The resulting area of the
         mosaic of images for each telescope is marked in black on an 
image of a plate plotted with Aladin.}
\label{map}
\end{figure*}

\begin{table}
\leavevmode
\caption {Log of the observations}
{\scriptsize
\begin {tabular} {ccccccccc}
\hline
   Telescope  & Date & Seeing($\arcsec$) & n. of frames & & Exp. & Times &($\sec$)&  \\
              &      &          &  & $u$ & $v$ & $b$ & $y$ & $H_{\beta}$ \\
\hline
  1.23 m CAHA  & 1999/01/12-15 & (1) & 19 & 1900 &  800 & 400 & 400 & 2000 \\

  1.23 m CAHA  & 2000/01/05-10 & 1.0-1.3 & 54 & 2200 & 1400 & 900 & 800 & 1400 \\

  1.52 m OAN   & 1999/01/13-16 & (1) & 36 & 1900 & 800 & 400 & 400 & 2000 \\

  1.52 m OAN   & 2000/02/07-14 & (1) & 23 & -- & -- & 900 & 800 & 1400 \\

   1 m JKT &  1998/12/11-14 & (1) & 81 & 2000 & 1200 & 800 & 700 & 1200 \\

  1 m JKT &  2000/02/02-06 & 0.9-1.3 & 37 & -- & -- & -- & -- & 2000 \\

  2.5 m WFC-INT & 2000/02/02-03 & 1.3 & 23 & 2000 & 2000 & 1200 & 500 & --  \\
\hline
\label{log}
\end {tabular}
(1) Poor weather conditions. Images not used in the final data.
}
\end {table}

\begin{table}
\leavevmode
\caption {Chip specifications}
{\scriptsize
\begin {tabular} {ccccccc}
\hline
    & 1.23 m CAHA & 1.52 m OAN & 2.5 m WFC-INT & 1 m JKT \\
\hline
     Type:          & SiTe2b & TK1024AB & 4$\times$ EEVi42-80 & SiTe2 \\
     Dimensions: &2048$\times$2048 & 1024$\times$1024 &4$\times$ 2048$\times$4100 & 2048$\times$2048 \\
     Pixel size:    & 24$\mu$=0.4$\arcsec$ & 24$\mu$=0.4$\arcsec$ & 13.5$\mu$=0.33$\arcsec$ &24$\mu$=0.33$\arcsec$ \\
     Field of view: & 10$^{\prime}$$\times$\-10$^{\prime}$ &6$^{\prime}$\llap{.}9$\times$6$^{\prime}$\llap{.}9 & $34.2$$^{\prime}$$\times$34.2$^{\prime}$ & 10$^{\prime}$\-$\times$\-10$^{\prime}$  \\
     Gain:           & 2.6e-/ADU & 6.55e-/ADU & 2.8 e-/ADU & 1.41e-/ADU \\
     Read-out noise: & 6.0 e- & 6.4 e- & 7.7 e- &7.5 e- \\
     Dymamic range: & 65553 & 65536  & 65000 & 65000 \\
     Typical bias level: & 154  & 470 & 1623 & 625 \\
     Overscan region: & right & top & right-top-left & top-right \\

\hline
\label{chips}
\end {tabular}
}
\end {table}

\end{document}